\documentclass[prd,onecolumn, nofootinbib, 11pt,]{revtex4}  
\usepackage{graphicx}
\usepackage{amsmath,braket}
\usepackage{amsfonts}
\usepackage{amssymb}
\usepackage{bm}
\usepackage{appendix}
\usepackage{mathtools}
\usepackage{comment}
\usepackage{bbold}
\usepackage{color}
\usepackage{slashed}
\usepackage[hyperindex=true, citecolor=green]{hyperref}
\usepackage{subfigure}
\usepackage{setspace}
\usepackage{enumitem}
\usepackage{longtable}
\usepackage{wasysym}
\usepackage[usenames,dvipsnames]{xcolor}
\usepackage{bm}
\usepackage{multirow}
\usepackage{changepage}
\usepackage{kantlipsum}
\usepackage{mathtools}
\usepackage[letterpaper, margin=.8in, top=0.85in, bottom=0.85in]{geometry}

\usepackage{tikz}
\usetikzlibrary{decorations.pathmorphing,arrows.meta,bending}

\pdfoutput=1

\newcommand{\Sch}{Schr\"{o}dinger }
\newcommand{\Dperp}{{\bf D}_\perp}

\DeclareMathAlphabet\mathbfcal{OMS}{cmsy}{b}{n}

\begin{document}

\singlespacing

\title{Conformal Structure of the Heavy Particle EFT Operator Basis}
\author{Andrew Kobach}
\affiliation{Physics Department, University of California, San Diego, La Jolla, CA 92093, USA}
\author{Sridip Pal}
\affiliation{Physics Department, University of California, San Diego, La Jolla, CA 92093, USA}

\date{\today}

\begin{abstract}
An operator basis of an effective theory with a heavy particle, subject to external gauge fields, is spanned by a particular kind of neutral scalar primary of the non-relativistic conformal group.  
We calculate the characters that can be used for generating the operators in a non-relativistic effective field theory, which accounts for redundancies from the equations of motion and integration by parts.  
\end{abstract}

\maketitle

\linespread{1}

\section{\normalsize Introduction}

If one can say that a particle, and not its antiparticle, exists in the laboratory, then the length scale of its spatial wave function $\Delta x$ is parametrically larger than its Compton wavelength $1/M$.   This hierarchy of scales leads to heavy particle effective field theories (heavy particle EFTs), where one can systematically include higher powers of $1/(\Delta x~ M)$.  Such systems are, in fact, fairly common.  For example, the $b$ quark can be located anywhere within a $B$ meson, which has a spatial size  of $\sim 1/\Lambda_\text{QCD}$, and heavy quark effective field theory is an expansion in powers of $\Lambda_\text{QCD}/m_b \sim 0.3$.     A more dramatic example is the electron in a hydrogen atom, whose wave function has a size $\Delta x \sim 10^{-10}$ m, and $1/(\Delta x ~ m_e) \sim 10^{-3}$, which is why the \Sch equation for single-partial quantum mechanics works so well in describing this system, using only the first-order expansion in $1/M$.  
Sometimes these theories are called non-relativistic effective field theories, insofar as there is a inertial frame in which there are non-relativistic particles.  

Even though heavy particle EFTs describe common physical scenarios, enumerating the independent operators that appear in the Lagrangian, i.e., defining the operator basis, takes considerable effort.  The reason for this is that defining the operator basis is more than just requiring that the operators preserve certain symmetries - it also involves accounting for non-trivial redundancies between  operators from the classical equations of motion and integration by parts~\cite{Politzer:1980me, Georgi:1991ch}.  There are popular EFTs with Lagrangians containing  heavy fields, e.g.,~NRQED (external abelian gauge fields), and HQET and NRQCD (external color gauge fields).  The operator basis for NRQED was written to order $\mathcal{O}(1/M^3)$ in Ref.~\cite{Kinoshita:1995mt} and to $\mathcal{O}(1/M^4)$ in Ref.~\cite{Hill:2012rh}.  The HQET/NRQCD operator basis was enumerated up to $\mathcal{O}(1/M^3)$ in Ref.~\cite{Manohar:1997qy}, and to $\mathcal{O}(1/M^4)$ by Ref.~\cite{Kobach:2017xkw}, which was later confirmed in Ref.~\cite{Gunawardana:2017zix}. 

A huge stride was taken recently by the authors of Refs.~\cite{Henning:2015daa, Henning:2015alf, Henning:2017fpj}, where they noticed that the operator basis for a relativistic EFT can be organized according to the representations of the conformal group.  In particular, accounting for the redundancies from the classical equations of motion can be mapped to the null conditions that saturate unitarity in the conformal group, and choosing the operator basis to be spanned only by primaries of the conformal algebra removes any redundancies associated with integration by parts.  By embedding operators into representations of the conformal group, one can use characters as inputs into a Hilbert series, which then can generate the operator basis, counting the number of operators in the EFT with the given field content.  Constructing the explicit operators with contracted internal and Lorentz indices, however, needs to be done by hand.  Even so, the Hilbert series output provides an invaluable tool for constructing a bonafide operator basis. 
For example, a Hilbert series aided in constructing the first correct operator basis for dimension-7 operators in the standard model EFT~\cite{Henning:2015alf} and dimension-8 operators in the HQET/NRQCD Lagrangian~\cite{Kobach:2017xkw}.

In practice, the characters of particular group representations are used as inputs for the Hilbert series, which generates all possible tensor products of these representations.  We refer the reader to Ref.~\cite{Lehman:2015via} for an introductory and pedagogical discussion regarding Hilbert series and the underlying group theory.
In Ref.~\cite{Kobach:2017xkw}, we constructed such a Hilbert series to help write down the operator basis for NRQED and HQET/NRQCD to order $\mathcal{O}(1/M^4)$.  
However, we did not use any organizational principle associated with the representation of the conformal group, since the equations of motion for non-relativistic fields are not those for relativistic fields.  
Instead, the ``characters" we used in the Hilbert series were constructed by hand. 
This begs the question:~{\it Are they characters of a group representation?}  Perhaps unsurprisingly, the answer is:~``{\it Yes}."  In this work, we show that the characters used in Ref.~\cite{Kobach:2017xkw} are those associated with ``shortened" representations of the non-relativistic conformal group, and the operator basis for heavy particle effective field theories are spanned by a special category of primary operators.  While this does have some analogy to the relativistic scenario studied in Refs.~\cite{Henning:2015daa, Henning:2015alf, Henning:2017fpj}, there are important subtleties with  non-relativistic theories, which we discuss in some detail.

\section{\normalsize Operator Basis for Heavy Particle EFTs}
\label{opbasis}

We consider operators that comprise an effective field theory that are singlets under the symmetries of the theory, each constructed out of the relevant degrees of freedom and any number of derivatives acting on them.  Of such operators, two or more may give rise to identical $S$-matrix elements, in which case they ought to not be counted as distinct.  This occurs when two or more operators:~(1) differ by a total derivative, or (2) can be related via the classical equations of motion (a kind of field redefinition)~\cite{Politzer:1980me, Georgi:1991ch}.   Accounting for these redundancies amounts to the program of constructing operator bases in effective field theories.  

How these redundancies apply to heavy-particle EFTs is described in Ref.~\cite{Kobach:2017xkw}, and we will briefly recapitulate it here.  Consider every possible rotationally- and gauge-invariant operator in the rest frame\footnote{In general, a covariant derivative $D_\mu$ can be written in terms of the velocity 4-vector $v^\mu$: $D_\mu = (v\cdot D) v^\mu + D_\perp^\mu$.  In the rest frame, i.e., where $v^\mu = (1,0,0,0)$, then $D_\mu = (D_t , \Dperp)$. } of a theory with only one heavy particle $\psi$:
\begin{equation}
\label{HQETLagrangian}
\mathcal{L} = \psi^\dagger i D_t \psi + \frac{1}{M^{d-1}}\sum_k^\infty c_k \psi^\dagger \mathcal{O}^{[d]}_k \psi \, ,
\end{equation}
where the $c_k$'s are dimensionless constants, $M$ is the mass of the heavy particle, and the operators $\mathcal{O}^{[d]}_k$ are Hermitian operators of mass dimension $d\geq 2$, constructed out of field strength tensors of gauge fields, covariant time derivatives, $D_t$, covariant spatial derivatives, ${\bf D}_\perp$ (and these derivatives can act to the right as well at the left), and spin vectors.  All covariant derivatives must be symmetric under exchange of spatial indices, otherwise they are proportional to field strength tensors, which have already been included.

Consider a set of operators that contain only one derivative.  For a given operator $\mathcal{O}$, the relationship between operators by integrating by parts is
\begin{eqnarray}
\psi^\dagger \mathcal{O} \partial \psi + [\partial \psi^\dagger ] \mathcal{O} \psi  + \psi^\dagger [\partial \mathcal{O} ] \psi &=&  \partial(\psi^\dagger \mathcal{O}  \psi ) = \text{total derivative}\, ,
\end{eqnarray}
where $\partial$ is a partial time or spatial derivative, and the square brackets indicate that the derivative only acts on the operator within the brackets.  
Combined with the identity
\begin{equation}
\psi^\dagger \mathcal{O} D \psi + [D \psi^\dagger ] \mathcal{O} \psi  + \psi^\dagger [D \mathcal{O} ] \psi =  D(\psi^\dagger \mathcal{O} \psi) \,,
\end{equation}
where $D$ is a covariant time or spatial derivative, and with the fact that $D (\psi^\dagger \mathcal{O} \psi) = \partial (\psi^\dagger \mathcal{O} \psi) $ since $\psi^\dagger \mathcal{O} \psi$ was defined to be a gauge singlet, we have the constraint:
\begin{equation}
\psi^\dagger \mathcal{O} D \psi + [D \psi^\dagger ]  \mathcal{O} \psi  + \psi^\dagger [D \mathcal{O} ] \psi =  \text{total derivative} \, . 
\end{equation}
This equation relates three operators to a total derivative, so to account for this redundancy, we need to ignore one of them.  One easy option is to only ignore operators $\mathcal{O}$ that contains derivative that act on $\psi^\dagger$.  But this solution cannot be generalized to the case with more derivatives in the operator, and accounting for the redundancies associated from integration by parts between operators with more  derivatives becomes  more challenging.

The equations of motion for $\psi$ also give relationship between operators:
\begin{equation}
i D_t \psi + \frac{1}{M^{d-1}}\sum_k^\infty c_k \mathcal{O}^{[d]}_k \psi = 0 \,  ,
\end{equation}
and multiplying from the left by $\psi^\dagger \mathcal{O}_j^{[d']}$:
\begin{equation}
\psi^\dagger \mathcal{O}_j^{[d']} i D_t \psi = - \frac{1}{M^{d-1}}\sum_k^\infty c_k \psi^\dagger \mathcal{O}_j^{[d']}\mathcal{O}^{[d]}_k \psi  \, .
\end{equation}
Any operator $\mathcal{O}$ that contains a covariant time derivative that acts on $\psi$ can be related to an infinite set of other operators at higher order.  
Therefore, this single equation that relates operators can be imposed if one ignores any operator that contains a covariant time derivative that acts on $\psi$.  
The same argument follows for $\psi^\dagger$. 

Lastly, there are relationships between operators due to the equations of motion of the field strength tensors associated with {\it external} gauge fields, i.e.,  $D_\mu F^{\mu\nu} = j^\nu$ and $D_\mu \widetilde{F}^{\mu\nu} = 0$, where $j^\mu = (\rho, {\bf J})$.  Because the effective theory defined in Eq.~\eqref{HQETLagrangian} is restricted to only the sector with one matter degree of freedom, there exists the possibility that whatever gauge fields appear in Lagrangian may have equations of motion that include external sources.  Representing the covariant derivative as $D_\mu = (D_t , {\bf D}_\perp)$ in the rest frame of the heavy particle, we have the non-abelian generalizations of Maxwell's equations:
\begin{eqnarray}
\Dperp \cdot {\bf E} &=& \rho  \, , \\
\Dperp \cdot {\bf B} &=& 0  \, , \\
\Dperp \times {\bf E} &=& -D_t {\bf B} \,  , \\
\Dperp \times {\bf B} &=& {\bf J } + D_t {\bf E}  \, .
\end{eqnarray}
So, if the operator $\mathcal{O}$ in Eq.~\eqref{HQETLagrangian} is constructed out of ${\bf E}$ and ${\bf B}$ (and covariant derivatives acting on them), then Maxwell's equations will make some operators vanish, as well as provide relationships between different operators.  Accounting for these constraints, one must impose $\Dperp \cdot {\bf B} = 0$, and choose to either express the operator $\mathcal{O}$ in terms of $\Dperp \times {\bf E}$ or $D_t {\bf B}$, but not both. 

In summary, imposing the constraints on operators from integration by parts and equations of motion of $\psi$, ${\bf E}$, and ${\bf B}$ on the rotationally- and gauge-invariant Lagrangian density in Eq.~\eqref{HQETLagrangian} will provide the operator basis for HQET.  That is, every operator gives rise to different $S$-matrix elements.  
There are additional symmetry constraints on such an EFT from residual relativistic boost symmetry~\cite{Luke:1992cs, Heinonen:2012km}.  This amounts to relating the coupling constants $c_k$ in Eq.~\eqref{HQETLagrangian} to one another, but does not alter the operator basis.

\section{\normalsize Operator Basis and the Schr\"{o}dinger Algebra}

We show that the characters in Ref.~\cite{Kobach:2017xkw}, which were constructed by hand in order to generate the operator basis for a heavy-particle effective theory, are, in fact, characters of irreducible representations of the non-relativistic conformal group (this group is also referred to as the \Sch group, and we will use these terms interchangeably).  Furthermore, from this one can determine that the operator basis for a heavy-particle effective field theory is spanned by particular kinds of primary operators of the non-relativistic conformal group.  
For those readers not familiar with symmetries of non-relativistic systems or the \Sch algebra, we invite them to read the Appendix, which is an introductory review to some of its well-known features that are relevant to the following discussion. 

The Lie algebra of the Schr\"{o}dinger group is:
\begin{equation}
\begin{gathered}
[J_i, J_j] = i \epsilon_{ijk} J_k\,, \quad \ {[J_i, K_j]} = i \epsilon_{ijk} K_k\,, \quad \ {[J_i, P_j]} =i \epsilon_{ijk} P_k\,, \\
{[H, K_i]} = - i P_i\,, \quad \ {[K_i, P_j]} = i N\delta_{ij}\,, \\
{[D, K_i]} =-i K_i\,, \quad \ {[D, P_i]} = iP_i\,, \quad \ {[D, H]} = 2iH\,, \\  
{[C, P_i]} = i K_i\,, \quad \ {[C, H]} = i D\,, \\  
{[K_i, K_j]} = {[H, P_i]} = {[H, J_i]} = {[P_i, P_j]} = {[N, \text{any}]} = 0\,, \\
  [D,J_i] = [C,J_i] = [C,K_i] = 0\, ,
\end{gathered} 
\end{equation}
where these are the generators of rotations ($J_i$), non-relativistic boosts ($K_i$), time translations ($H$), spatial translations ($P_i$), scaling transformations ($D$), special conformal transformations ($C$), and number charge ($N$). 
States that transform as irreducible representations of the \Sch algebra can be labeled with the eigenvalues of the Cartan generators for the group, i.e., $D$, $N$, and $J_3$:
\begin{eqnarray}
D\ket{\Delta, n,  m} &=& i\Delta \ket{\Delta, n,  m} \, , \\
N\ket{\Delta, n,  m} &=& n \ket{\Delta, n,  m} \, , \\
J_3\ket{\Delta, n, m} &=& m \ket{\Delta, n,  m} \, .
\end{eqnarray}
%
The \Sch algebra has raising and lowering operators, analogous to those for angular momentum, which raise and lower the scaling dimension $\Delta$:
\begin{eqnarray}
DP_i \ket{\Delta, n,  m} &=& i(\Delta+1)P_i \ket{\Delta, n,  m}\, ,  \\
DK_i \ket{\Delta, n,  m} &=& i(\Delta-1)K_i \ket{\Delta, n,  m}\, , \\
DH \ket{\Delta, n, m} &=& i(\Delta+2)H\ket{\Delta, n,  m}\, , \\
DC \ket{\Delta, n,  m} &=& i(\Delta-2)C\ket{\Delta, n,  m}\, .
\end{eqnarray}
$P_i$ and $H$ raise the scaling dimension, and $K_i$ and $C$ lower it.   Using the linear combinations:
\begin{equation}
P_{\pm} \equiv P_1\pm i P_2 \, , \hspace{0.25in} K_\pm \equiv K_1\pm i K_2 \, ,
\end{equation}
the following eigenvalue equations following directly from the algebra:
\begin{eqnarray}
J_3P_\pm \ket{\Delta, n,  m} &=& (m\pm1)P_\pm\ket{\Delta, n,  m}\, ,  \\
J_3P_3 \ket{\Delta, n,  m} &=& m P_3\ket{\Delta, n,  m}\,,\\
J_3K_\pm \ket{\Delta, n,  m} &=& (m\pm1)K_\pm\ket{\Delta, n,  m}\, , \\
J_3K_3 \ket{\Delta, n,  m} &=& m K_3\ket{\Delta, n,  m}\,,\\
J_3H \ket{\Delta, n,  m} &=& mH\ket{\Delta, n,  m}\, , \\
J_3C \ket{\Delta, n,  m} &=& mC\ket{\Delta, n,  m}\, .
\end{eqnarray}
More such relations exist, but we only list the ones here that we will use.  
The Cartan generator $N$ commutes with everything, and the action of other generators on the state does not change its number-charge. This allows us to consider states sector-wise depending on its number charge. 
In the sector of states with number charge $n\neq 0$, one can lower the scaling dimension by action of $K_i$ and $C$, but the unitarity bound restricts the lowest possible dimension. For details, see Appendix~\ref{app3}.  There can be lowest-weight states of scaling dimension (in group theory literature, this is known as highest-weight state), such that:
\begin{eqnarray}
K_i \ket{\Delta_{*}, n,  m} &=& 0  \, , \\
C \ket{\Delta_{*}, n, m} &=& 0  \, ,
\end{eqnarray}
where $m=-j,-j+1,\cdots, j$ where $j$ is the total spin of the highest-weight state. We note that even though the highest-weight states can be assigned a total spin $j$, this is no longer true once we act on these states by $P_\pm$ or $P_3$, since acting on a state with spin $j$, they produce a linear combination of spin $j+1,j,\cdots , |j-1|$. 
If the state's scaling dimension is $\Delta_{*} > d/2$, where $d$ is the number of spatial dimensions, then an irreducible representation of the \Sch algebra can generated by acting repeatedly on $\ket{\Delta_{*}, n,  m}$ with $P_i$ and $H$.  Thus all the states in the representation are of the form:
\begin{align}
\ket{\Delta, n,  m^\prime}=H^\ell P_+^r P_-^p P_3^q \ket{\Delta_*, n,  m} \, ,
\end{align}
where, specifically, $\Delta=2\ell+r+p+q+\Delta_{*}$, $m^{\prime}=r-p+m$ and $m=-j,-j+1,\cdots, j$.
The character for this representation is a trace over all its states (following the procedure for relativistic conformal representation, as detailed in Ref.~\cite{Barabanschikov:2005ri}):
\begin{eqnarray}
\chi_{[\Delta > d/2, n\neq 0, j]} &=& \text{Tr} \left[ e^{i\theta_D D + i\theta_N N + i\theta_3 J_3 }\right] , \\
\label{charsum}
&=& e^{in\theta_N } \displaystyle \sum_{\substack{ |m| \leq j \\ \ell, r, p, q \geq 0 }}  \bra{\text{adjoint}} e^{i\theta_D D  + i\theta_3 J_3} H^\ell P_+^r P_-^p P_3^q \ket{\Delta_*, n, m} , \\
\label{longrep}
&=& \frac{e^{in\theta_N } ~\Lambda^{\Delta_{*}}~\chi^{SU(2)}_{(j)}(z)}{(1-z^2\Lambda ) (1-\Lambda ) (1-\Lambda/z^2 ) (1-\Lambda^2 )} \, ,
\end{eqnarray}
where $\Lambda \equiv e^{-\theta_D}$, $z  \equiv e^{i\theta_3/2}$, and $ \bra{\text{adjoint}}$ means the complex conjugate of the state of $H^\ell P_+^r P_-^p P_3^q \ket{\Delta_*, n, m}$, such that the norm of the state is unity.  Here, $\chi^{SU(2)}_{(j)}(z)$ is the character for an $SU(2)$ $j-$plet, i.e.,
\begin{equation}
\chi^{SU(2)}_{(j)}(z) \equiv \sum_{|m|\leq j} \bra{j, m} z^{2J_3} \ket{j,m} . 
\end{equation}
For example, the character for an $SU(2)$ doublet is $\chi^{SU(2)}_{\bf 2} = z + 1/z$, and the character for an $SU(2)$ triplet is $\chi^{SU(2)}_{\bf 3} = z^2 + 1 + 1/z^2$, and so on.  Since $P_i$ and $H$ are the generators for spatial and time translations, respectively, we can identify the term $[(1-z^2\Lambda ) (1-\Lambda ) (1-\Lambda/z^2 )]^{-1}$ as the generating functional for all possible {\it symmetric} products of spatial derivatives, and $(1-\Lambda^2 )^{-1}$ as the generating functional for all possible products of time derivatives.  To make this more clear, we can put in the numbers $\mathcal{D}_t$ and $\mathcal{D}_\perp$ (of modulus less than unity) to flag where and how many derivatives are generated, e.g., 
\begin{eqnarray} 
P_0(\Lambda) \equiv \frac{1}{(1-\Lambda^2\mathcal{D}_t )} &=& 1 + \Lambda^2 \mathcal{D}_t + \Lambda^4\mathcal{D}_t^2 + ... \\ 
P_\perp(\Lambda) \equiv \frac{1}{(1-z^2\Lambda\mathcal{D}_\perp) (1-\Lambda\mathcal{D}_\perp)(1-\Lambda\mathcal{D}_\perp/z^2)} &=& 1 + \Lambda\mathcal{D}_\perp \chi^{SU(2)}_{\bf 3} + \Lambda^2\mathcal{D}_\perp^2\left( 1 + \chi^{SU(2)}_{\bf 5} \right) + ... \nonumber \\
\end{eqnarray}
We can illustrate the behavior of these generating functionals with an example.  Consider that the generating function for spatial derivatives acts on an object that is a singlet under rotation, call it $\phi$, then the rotational indices can be reintroduced by hand, and generating derivatives can be represented as: $P_\perp \phi = \phi + \partial_i \phi + \partial_i\partial_i \phi + \partial_i \partial_j \phi + \cdots$.   Note that there is no term like $\epsilon_{ijk} \partial_i \partial_j \phi$ generated; it is trivially zero.

If the scaling dimension of the highest-weight state $\ket{\Delta_*, n, m}$ in the representation is $\Delta_* = d/2$, then the unitarity bound is saturated, leading to the fact that the following state has zero norm (see Appendix~\ref{app3} and Refs.~\cite{Nishida:2007pj, Goldberger:2014hca, Pal:2018idc}):
\begin{equation}
\left( H - \frac{P_i^2}{2n} \right) \ket{\Delta_*=d/2, n, m} = 0 \, .
\end{equation}
Therefore, the character for the representation when the highest-weight state has $\Delta_*= d/2$ should not contain the contribution coming from the state $\left( H - \frac{P_i^2}{2n} \right) \ket{\Delta_*=d/2, n, m}$ and any power of $H$ or $P_i$ acting on it.  This can be achieved by removing the tower of states generated by $H$ acting on $\ket{\Delta_*=d/2, n, m}$.  As discussed in Section~\ref{opbasis}, this is precisely the requirement that when defining an operator basis with a heavy particle, taking into account the equations of motion, that one can choose a basis with no time derivatives act on heavy field $\psi$. 
The character for such a shortened representation is can be easily calculated:
\begin{equation}
\chi_{[\Delta_* = d/2, n\neq 0, j]} = e^{in\theta_N } ~\Lambda^{d/2}~P_\perp(\Lambda)~\chi^{SU(2)}_{(j)}(z)  \,  .
\end{equation} 
This is the character used in Ref.~\cite{Kobach:2017xkw} for the heavy particle degree of freedom, modulo the multiplicative factor of $e^{in\theta_N } ~\Lambda^{d/2}$.   Therefore, for the sake of defining an operator basis, one can say that the heavy-particle state is a highest-weight state with scaling dimension $\Delta_* = d/2$ and $n\neq 0$.  And, in particular, if it is a heavy fermion, then $\chi^{SU(2)}_{(j)} = \chi^{SU(2)}_{\bf{2}}$. This is a scenario when the equation of motion can be derived using the algebra and the constraint from  unitarity, though this does not always have to the case.\footnote{To cite a specific example, at the interacting fixed point of a relativistic $\phi^4$ theory, the equation of motion is not associated with a unitarity bound, since, in $4-\epsilon$ dimensions, the field $\phi$ acquires an anomalous dimension,  which no longer saturates the unitarity bound~\cite{Rychkov:2015naa}.}  
The representation for $\psi^\dagger$ is the same as $\psi$, but with the sign of $n$ flipped.

In the heavy particle EFT, the Lagrangian is expressed using {\it external} electric and magnetic fields (or their non-abelian generalizations).  We are interested in embedding these fields within an representation of the \Sch group, where they would have well-defined charges under scaling transformations, number charge, and $z$-component of angular momentum.  To begin, one must take care to reinstate the location of the speed of light constant $c$.  Because space time scale differently under scaling transformations, i.e., $x \rightarrow \lambda x$ and $t \rightarrow \lambda^2 t$, the speed of light is not invariant, behaving as an intrinsic scale in the  theory.  As such, $\mathbfcal{E} \equiv {\bf E}/c$ and ${\bf B}$ are the fundamental fields that appear in the field strength tensor $F^{\mu\nu}$.   Since they are externally defined, they can be taken to scale in a similar way, both with scaling dimension $\Delta = 2$.  Also, $\mathbfcal{E}$ and ${\bf B}$ transform as vectors under rotation, so they are both spin-1.   Lastly, since the electric and magnetic fields are Hermitian, they can not carry any number-charge, so they have $n=0$.  As noted in Appendix~\ref{app2}, the representation of the \Sch group for operators with $n=0$ differs from the $n\neq 0$ sector.  For example, consider a highest-weight state $\ket{\Delta_*, n=0, m}$ such that $K_i\ket{\Delta_*, n=0, m} =C\ket{\Delta_*, n=0, m} = 0$.  The \Sch algebra then leads to the following:
\begin{eqnarray}
K_j P_i \ket{\Delta_*, n=0, m} &=& 0 \, , \\
C P_i \ket{\Delta_*, n=0, m} &=& 0 \, .
\end{eqnarray}
Therefore, the state $P_i \ket{\Delta_*, n=0, m}$ is also a highest-weight state in scaling dimension.  In order to embed $\mathbfcal{E}$ and ${\bf B}$ in the \Sch representation, we can choose to define the following kind of state $\ket{\Delta_*, n=0, m}$, where:~(1) $C \ket{\Delta_*, n=0, m} = 0$, and (2) $\ket{\Delta_*, n=0, m} \neq P_i \ket{\Delta'_*, n=0, m'}$, where $\ket{\Delta'_*, n=0, m'}$ is some other state in the Hilbert space.  This is the definition of the state in the \Sch group that we associate with the electric and magnetic fields.  As before, one can build up a representation of the \Sch group (two towers of states) by acting by $P_i$ and $H$ on this $\ket{\Delta_*, n=0, m}$.  In $d=3$ spatial dimensions there is no constraint from unitarity regarding how high these towers can go.  However, these towers do not extend forever, since two of Maxwell's equations are:  
\begin{eqnarray}
\label{max1}
\Dperp \cdot {\bf B} = 0\,,\ \hspace{0.25in} \Dperp \times \mathbfcal{E} = -\frac{1}{c}D_t {\bf B}\, . 
\end{eqnarray}
The other two Maxwell's equation with source term do not constrain or relate the tower of states, since both the current and charge density are externally defined.  If we take the $c\rightarrow \infty$ limit, then Eqs.~\eqref{max1} gets contracted from the Poincar\'{e} representation to the $N=0$  representation of the Galilean group (for some details regarding this $c\rightarrow \infty$ contraction, see Appendix~\ref{app1}), and end up being invariant under scaling and special conformal transformations:
\begin{eqnarray}
\label{max2}
\Dperp \cdot {\bf B} = 0\,,\ \hspace{0.25in} \Dperp \times \mathbfcal{E} =  0 \, . 
\end{eqnarray}
Therefore, these are the shortening conditions for the states in the \Sch group associated with the electric and magnetic fields.  So, the  the characters for $\mathbfcal{E}$ and ${\bf B}$ are:
\begin{eqnarray}
\chi^\mathcal{E}_{[\Delta_* = 2, n=0, j=1]} &=& \Lambda^2 ~P_0(\Lambda) ~P_\perp(\Lambda)~\bigg( \chi^{SU(2)}_{\bf 3} - \Lambda \mathcal{D}_\perp\chi^{SU(2)}_{\bf 3}  + \Lambda^2 \mathcal{D}_\perp^2 \bigg) \, , \\
\chi^B_{[\Delta_* = 2, n=0, j=1]} &=& \Lambda^2 ~P_0(\Lambda) ~P_\perp(\Lambda)~\bigg( \chi^{SU(2)}_{\bf 3} - \Lambda \mathcal{D}_\perp \bigg) \, .
\end{eqnarray}
The additional $\Lambda^2 \mathcal{D}_\perp^2$ term in the character for $\mathbfcal{E}$ is due to the fact that if one subtracts out $\Lambda\mathcal{D}_\perp \chi^{SU(2)}_{\bf 3}$, then one will also subtract out $\Lambda^2\mathcal{D}_\perp^2$, but this term was never there to begin with, since derivatives are symmetric under interchange of their spatial indices.  These are precisely the characters used in Ref.~\cite{Kobach:2017xkw} for the external gauge fields.

We have established how the requirements of defining an operator basis for a heavy particle EFT can be associated with the relevant degrees of freedom in the theory, i.e., the heavy particles, electric and magnetic fields, and the time and spatial derivatives that act on them, falling into irreducible representations of the \Sch group.  In particular, we have shown that characters of certain representations of the \Sch group match those we used in Ref.~\cite{Kobach:2017xkw}, where we only had in mind the constraints from the equations of motion.  At last, we can take tensor products between these representations of the \Sch group to generate operators that appear in the Lagrangian.  Illustrating this with the following cartoon (including the details of the shortening conditions for $\mathbfcal{E}$ and ${\bf B}$ is a bit cumbersome):
\begin{equation}
\label{tensorprod}
\left( \begin{array}{c} \psi^\dagger \\ \mathcal{D}_\perp \psi^\dagger \\ \mathcal{D}^2_\perp \psi^\dagger \\ \vdots \end{array} \right) \otimes
\left( \begin{array}{c} \mathbfcal{E} \\ \mathcal{D}_\perp \mathbfcal{E} \\ \mathcal{D}^2_\perp \mathbfcal{E}, \mathcal{D}_t \mathbfcal{E} \\ \vdots \end{array} \right) \otimes 
\left( \begin{array}{c} {\bf B} \\ \mathcal{D}_\perp {\bf B} \\ \mathcal{D}^2_\perp {\bf B}, \mathcal{D}_t {\bf B} \\ \vdots \end{array} \right) \otimes 
\cdots \otimes 
\left( \begin{array}{c} \psi \\ \mathcal{D}_\perp \psi \\ \mathcal{D}^2_\perp \psi \\ \vdots \end{array} \right)
~=~ \left( \begin{array}{c} \mathcal{O} \\ \mathcal{D}_\perp \mathcal{O} \\ \mathcal{D}^2_\perp \mathcal{O}, \mathcal{D}_t \mathcal{O} \\ \vdots \end{array} \right) + \cdots 
\end{equation}
Because the right-hand side of Eq.~\eqref{tensorprod} are also representations of the \Sch group, only the highest-weight operators are not associated with total derivatives.  These highest-weight operators have the properties that $[C, \mathcal{O}]= 0$, and $ \mathcal{O} \neq [P_i, \mathcal{O}']$, where $\mathcal{O}'$ is some other operator in the Hilbert space.  Therefore, it is exactly these highest-weight operators that span the operator basis for the EFT with one heavy particle.  This connection between representations of the \Sch group and a heavy particle EFT can be made for EFTs with multiple heavy fields.   If there are no interactions that cause heavy fields to transform into other types, then each heavy field can be labeled with a different charge $n$, and operator basis is required be invariant under $N$.

\section{\normalsize Discussion and Conclusions}

The operator basis for a heavy particle EFT, subject to external gauge fields, can be organized according to the representations of the non-relativistic conformal group (often called the \Sch group).  Such an organization allows one to easily remove any redundancy between operators due to integration by parts and the equations of motion for the individual degrees of freedom.  Specifically, we discuss that the heavy particle states are highest-weight states that saturate unitarity in $d=3$ spatial dimensions, and this leads to a representation of the \Sch group that removes any time derivatives acting on the heavy field, which amounts to the same imposition on the operator basis due to the equations of motion of the heavy field.  The external gauge fields are associated with the $n=0$ sector of the \Sch group, and this necessitates an extended classification of highest-weight states~\cite{Pal:2018idc}.  Maxwell's equations in the $c\rightarrow \infty$ limit produce the shortening conditions for the external gauge fields.  Taken together, the tensor products of these \Sch representations is itself a representation of the \Sch group, and the highest-weight (with $n=0$ and $j=0$) operators of this are exactly those which are not total derivatives, and therefore are the operators that span the operator basis for an EFT with a heavy particle.  

We have shown the characters of representation for the neutral sector and charged sector of the \Sch group to be precisely those used in Ref.~\cite{Kobach:2017xkw}, which used a Hilbert series to help tabulate the operator basis for NRQED and HQET/NRQCD, up to and including operators at $\mathcal{O}(1/M^4)$.  
An analogous situation occurs in relativistic theories, and the authors of Refs.~\cite{Henning:2015daa, Henning:2015alf, Henning:2017fpj} discuss why it may not be unreasonable to intuit a connection between an operator basis and the representations of a conformal group.  It is also worth mentioning that since the \Sch group is non-compact, it has similar subtleties associated with character orthogonality as in the relativistic conformal group. Operationally, this does not hamper the operator counting program, nonetheless it would be interesting to explore further the subtleties associated with character orthogonality of Sch\"odinger algebra in the same spirit as done in relativistic case.

The methodology introduced here could be leveraged to also describe the operator content beyond the realm of heavy particle effective field theory with a single heavy field, for example, fermions at unitarity \cite{Zwierlein:2004zz,Regal:2004zza}, two nucleon systems \cite{Kaplan:1998we,Kaplan:1998tg}, and it could have potential application in writing an operator basis for anisotropic Weyl anomaly, i.e., the anomaly associated with nonrelativistic scaling upon coupling the theory with curved space-time~\cite{Arav:2016xjc,Pal:2016rpz,Auzzi:2016lrq}.

\acknowledgements{We are appreciative for discussions with Jaewon Song, and we thank Ken Intriligator, Aneesh Manhoar, and John McGreevy for comments and feedback.  This work is supported in part by DOE grant \#DE-SC0009919.}

\appendix

\section{\normalsize Symmetries of Non-Relativistic Systems}
\label{app1}
In this appendix, we review the symmetries associated with non-relativistic systems.  
Much of the details that we present below can be found in, for example, Refs.~\cite{levy1971galilei,Niederer:1972zz,Henkel:1993sg,Mehen:1999nd,Henkel:2003pu,Nishida:2007pj,Goldberger:2014hca,Golkar:2014mwa,Pal:2018idc}.
Newton's equation of motion 
for a particle with mass $m$ subject to an external force ${\bf F}$ at time $t$ and position ${\bf x}$ is:
\begin{equation}
\label{newton}
{\bf F} (t, {\bf x}) = m \frac{d^2 {\bf x}}{\ \!dt^2} \, .
\end{equation}
Consider a change in time and space coordinates $(t,{\bf x})$ to $(t',{\bf x}')$ defined by the following transformations:
\begin{equation}
\label{galileantrans}
t \mapsto t' = t + b\,, \hspace{0.5in} x_i \mapsto x_i' = R_{ij}x_j + v_i t + a_i\,.
\end{equation}
Here, $R_{ij}$ is a rotation matrix, $v_i$ is the velocity, $b$ is a translation in time, and $a_i$ is a translation in space. Roman letters indicate space indices, and $v_i$, $b$, and $a_i$ are all real constants, independent of time.  Eq.~\eqref{newton} becomes the following under such a transformation:
\begin{equation}
F_i(t, {\bf x}) = m \frac{d^2 x_i}{\ \!dt^2} ~~~\mapsto ~~~ \widetilde{F}_i(t', {\bf x}') \equiv R_{ij}{F}_j(t, {\bf x}) = m R_{ij}\frac{d^2 x_j}{dt^2}=m \frac{d^2 x_i'}{\ \!dt'^2}\,  ,
\end{equation}
which implies 
\begin{equation}
\widetilde{F}_i(t',{\bf x}') = m\frac{d^2 x'_i}{\ \!dt'^2}\,  .
\end{equation}
The transformations defined in Eq.~\eqref{galileantrans} leave the {\it form} of Eq.~\eqref{newton} unchanged and are, therefore, symmetries associated with that equation of motion.  The transformations in Eq.~\eqref{galileantrans} are the most general {\it linear} transformations that leave time and space intervals (defined at the same moment in time) separately unchanged, i.e., 
\begin{equation}
t_1 - t_2 = \text{const}, \hspace{0.5in} |{\bf x}_1 - {\bf x}_2 | = \text{const, if $t_1 = t_2$}. 
\end{equation}
These transformations furnish what is known as the {\it inhomogeneous Galilean group}.  We refer to the inhomogeneous Galilean group as the Galilean group when no confusion is likely.  The group multiplication laws and inverses of group elements are straightforward to work out. 

The elements of the Galilean group can be represented as $\exp[i\theta_a X_a]$, where $\theta_a$'s are the 10 parameters needed to define a Galilean transformation, and the $X_a$'s are the generators of the group.  The following commutation relations define the Lie algebra:
\begin{eqnarray}
\begin{gathered}
[J_i, J_j] = i \epsilon_{ijk} J_k\,, \quad \ {[J_i, K_j]} = i \epsilon_{ijk} K_k\,,  \quad \ {[J_i, P_j]} = i \epsilon_{ijk} P_k\,, \\
 {[H, K_i]} = - i P_i\,,  \quad {[K_i, P_j]} = 0\,, \\
{[K_i, K_j]} = {[H, P_i]} = {[H, J_i]} = {[P_i, P_j]}  = 0\, , 
\end{gathered} 
\end{eqnarray}
where $P_i$ generates spatial translations, $H$ generates time translations, $J_i$ generates rotations, and $K_i$ are generators of Galilean boosts (named as such to distinguish them from Lorentz boosts).  Here, we use the convention that the antisymmetric tensor $\epsilon_{123} = 1$.  Note that the specific commutation relations $[K_i, K_j]$ and $[P_i, K_j]$ amount to the only differences compared to the Lie algebra of the relativistic Poincar\'{e} group.

The Lie algebra of the Galilean algebra can also be derived from the Poincar\'{e} algebra by reintroducing the speed of light,  i.e.,  $H\rightarrow H/c$, $K_i \rightarrow cK_i$, and letting $c\rightarrow \infty$. For example, the generators of Lorentz boosts have the commutation relation $[K_i, K_j] = - i\epsilon_{ijk}J_k$.  Putting the factors of $c$ back in: $[K_i, K_j] = - i\epsilon_{ijk}J_k/c^2$, and in the limit $c\rightarrow \infty$, one induces the commutation relation for Galilean boosts: $[K_i, K_j] = 0$.  Likewise, for the commutator $[K_i, P_j] = iH\delta_{ij} \rightarrow [K_i, P_j] = iH\delta_{ij}/c^2$ in the Poincar\'{e} group, this relation becomes $[K_i, P_j]=0$ in the Galilean group,  i.e., when $c\rightarrow \infty$. 

There exists the possibility that the Galilean group can be augmented with an additional generator, called $N$, such that $N$ commutes with all other generators:
\begin{equation}
{[N,\text{any}]} = 0\, ,
\end{equation}
and the commutation relation ${[K_i, P_j]}$ becomes:
\begin{equation}
{[K_i, P_j]} = 0 \hspace{0.25in} \longrightarrow \hspace{0.25in} {[K_i, P_j]} = iN\delta_{ij} \, .
\end{equation}
The augmentation of this kind is known as \textit{central extension}\footnote{Originally, the algebra without the central extension was called by Galilean algebra.  Lately, the trend in the literature is to call the centrally-extended algebra the ``Galilean algebra," and specify $N=0$ as a special case. }~\cite{levy1971galilei}.  When taking the $c\rightarrow \infty$ limit in the Poincar\'{e} algebra, one obtains the $N=0$,  i.e., the chargeless (neutral) sector of the Galilean group, which is its own algebra.

It is interesting to consider the special case when ${\bf F}(t,{\bf x}) = 0$.  Here, the classical equation of motion is:
\begin{equation}
\label{freeeom}
\frac{d^2{\bf x}}{\ \!dt^2} = 0\,.
\end{equation}
This is invariant under the Galilean transformations.  In addition to the transformations contained in the Galilean group, one could consider a kind of scaling transformation that acts like:
\begin{eqnarray}
\label{scaling}
x_i \mapsto x_i' = \lambda x_i \, , \hspace{0.5in} t \mapsto t' = \lambda^z t \, .
\end{eqnarray}
For any value of $z$, the classical equation of motion is invariant under such transformations. However, if the classical action,  which appears as $\exp(i S)$ in path integral, 
\begin{equation}
\label{freeS}
S = \int dt \left[\displaystyle\sum_i \frac{1}{2}  m \left(\frac{dx_i}{dt}\right)^2  \right] \, ,
\end{equation}
is  to be invariant under such transformations, then it is necessary that $z=2$. This difference in the requirements of the value of $z$ illustrates the important point that the equation of motion, being associated with the extremum of the action, is less constraining than the action, since the latter contains information regarding all possible configurations of the system in time. 
There is one additional kind of transformation which leaves both the classical equation of motion and the classical action invariant:
\begin{eqnarray}
\label{sct}
x_i \mapsto x_i' = \frac{x_i}{(1+kt)}\, , \hspace{0.5in} t \mapsto t' = \frac{t}{(1+kt)}\, , 
\end{eqnarray}
where  $k$ is a real number.  These transformations are known as the {\it special conformal transformations}.  Together, the Galilean transformations, the scaling transformations in Eq.~\eqref{scaling} (when $z=2$) and the special conformal transformations form what is called the {\it Schr\"{o}dinger group},\footnote{The group was popularized in the context of the free Schr\"{o}dinger equation~\cite{Niederer:1972zz}, hence the name.  It does not necessarily have to do with quantum mechanics.   }  where temporal and spatial coordinates transform in the following way:
\begin{equation}
t \mapsto t' = \lambda^2 \left(\frac{t+b}{1+k(t+b)} \right), \hspace{0.5in} x_i \mapsto x_i' = \lambda \left( \frac{R_{ij}x_j + v_i t + a_i}{1+k(t+b)} \right) .
\end{equation}
The group multiplication laws and inverses are straightforward  (albeit algebraically tedious) to calculate.\footnote{We note that the transformation of the time coordinate \cite{Goldberger:2014hca,Pal:2018idc} can be mapped on to a SL(2,$\mathbb{R}$) group,  i.e., transformations of the form 
\begin{equation}
t \mapsto t' = \frac{at +b}{ct+d}\,, \hspace{0.25in} ad - bc = 1\,.
\end{equation}  In terms of these parameters, the inverse and the group multiplication is relatively less tedious to compute.}

The elements of the \Sch group can be represented as $\exp[i\theta_a X_a]$, where in the index $a$ runs over the number of generators.  The Lie algebra of the Schr\"{o}dinger group is:\footnote{These commutation relations are the same as those in Refs.~\cite{Nishida:2007pj, Goldberger:2014hca}, but differ from those in Ref.~\cite{Niederer:1972zz} by some sign conventions.}

\begin{equation}
\begin{gathered}
[J_i, J_j] = i \epsilon_{ijk} J_k\,, \quad \ {[J_i, K_j]} = i \epsilon_{ijk} K_k\,, \quad \ {[J_i, P_j]} =i \epsilon_{ijk} P_k\,, \\
{[H, K_i]}  = - i P_i\,, \quad \ {[K_i, P_j]} = i N\delta_{ij}\,, \\
{[D, K_i]} =-i K_i\,, \quad  \ {[D, P_i]} = iP_i\,, \quad \ {[D, H]} = 2iH\,,\quad \ {[D, C]} = -2iC\,, \\  
{[C, P_i]}  = i K_i\,, \quad \ {[C, H]} = i D\,, \\  
{[K_i, K_j]} = {[H, P_i]} =  {[H, J_i]} = {[P_i, P_j]} = {[N, \text{any}]} = 0\,, \\
  [D,J_i] = [C,J_i] = [C,K_i] = 0\, ,
\end{gathered} 
\end{equation}
where $D$ is the generator of scaling transformations, and $C$ is the generator of the special conformal transformations.  
The Cartan generators for the \Sch group are $E_{1}\equiv-i D$, $E_{2}\equiv J_3$, and $E_{3}\equiv N$,  i.e., the maximally commuting set of generators. The generators with definite weight under these Cartan generators are given by:
\begin{align}
J_\pm &\equiv J_1 \pm i J_2\, , \\
P_\pm &\equiv P_1 \pm i P_2\,,\\
K_\pm &\equiv K_1 \pm i K_2\,,\\
P_3\,, &~H\,\!,  ~K_3\,,~ C\, .
\end{align} 
A generator $X$ carries a weight $w$ under a Cartan generator $E$ if $[E,X]=wX$.  The factor of $-i$ with generator $D$ makes the weights real.  The weights follow directly from the algebra, and are tabulated in Table~\ref{weighttable}.  
\begin{center}
\begin{table}[!ht]
\begin{tabular} {|c||c|c|c|}
\hline
& $E_{1}\equiv-i D$ & $E_{2}\equiv J_3$& $E_{3}\equiv N$\\
 \hline\hline
$P_{\pm}$ & $1$ &\ $\pm1$&$0$\\
 \hline
$P_3$ & $1$& $\ 0$ &$0$\\
 \hline
$H$ & $2$&$\ 0$ & $0$\\
\hline
$J_{\pm}$ & $0$&$\pm1$ & $0$\\
 \hline
$K_{\pm}$ & $\!\!\!-1$ & $\pm1$&$0$\\
 \hline
$K_{3}$ & $\!\!\!-1$ & $\ 0$&$0$\\
\hline
$C$ & $\!\!\!-2$ & $\ 0$&$0$\\
\hline
\end{tabular}
\caption{Table for weights of generators $P_\pm,~P_3,~H,~J_\pm,~K_\pm,~K_3,~C$ under the Cartan generators $E_{1}\equiv-i D$, $E_{2}\equiv J_3$, and $E_{3}\equiv N$.  The $ij$-th entry is the weight of the $i$'th generator ($i$ running over the possibilities $P_\pm,~P_3,~H,~J_\pm,~K_\pm,~K_3,~C$ under the Cartan generator $E_j$.}
\label{weighttable}
\end{table}
\end{center}
%

\section{Symmetries, Operators, and States}
\label{app2}

If $U$ is a group element, then assume there exists a Hilbert space with a vacuum state such that all $U$'s leave the vacuum invariant:
\begin{eqnarray}
U \ket{0} = \ket{0} \, .
\end{eqnarray} 
If $U$ can be represented as $\exp[i\theta_a X_a]$, where $X_a$ are the generators of the group's Lie algebra, then all $X$'s annihilate the vacuum:
\begin{equation}
X_a \ket{0} = 0 \, .
\end{equation}
Consider that the Hilbert space is spanned by more states $\ket{\mathcal{O}_A}$ (where $A$ is just an arbitrary label) than just the vacuum, which are defined as local operators $\mathcal{O}_A(\tau,{\bf x})$ acting on the vacuum where $\tau=i t$.\footnote{The states are prepared in Euclidean time $\tau$ to ensure finite norm as we will see later.}  For now, let us only consider (gauge invariant) states, created by operators acting at the origin:
\begin{equation}
 \mathcal{O}_A(0,{\bf 0}) \ket{0} =\ket{\mathcal{O}_A}   \, .
\end{equation}
Because acting on a state with a group transformation produces, in general, a linear combination of states still within that Hilbert space, one can say acting with a generator  produces a linear combination of states:
\begin{equation}
X \ket{\mathcal{O}_A} = L_{AB} \ket{\mathcal{O}_B} \, , 
\end{equation}
where $L$ is some matrix, which depends on what $X$ was chosen.   All the generators annihilate the vacuum, therefore
\begin{equation}
X \ket{\mathcal{O}_A} = [X,\mathcal{O}_A(0,{\bf 0})] \ket{0} = L_{AB} \mathcal{O}_B(0,{\bf 0}) \ket{0} \,  .
\end{equation}

Consider that the Hilbert space is spanned by operators that transform non-trivially under the \Sch group transformations.  The largest set of Cartan generators is $-i D$, $N$, and $J_3$, and we can always choose to label our states according the eigenvalues of these operators:
\begin{eqnarray}
 D \ket{\Delta, n,  m} &=&  i\Delta \ket{\Delta, n,  m}\, , \\
N \ket{\Delta, n,   m} &=& n \ket{\Delta, n,  m}\, , \\
J_3 \ket{\Delta, n,  m} &=&  m \ket{\Delta, n,  m} \, .
\end{eqnarray}
The eigenvalue of $D$ is complex, but we can see the significance of this in terms of the transformation properties of operators.   Say that there exists an operator $\mathcal{O}_{[\Delta, n, m]}$ that creates the state $\ket{\Delta, n,  m}$ at the origin:
\begin{equation}
\ket{\Delta, n,  m} \equiv \mathcal{O}_{[\Delta, n, m]}(0,{\bf 0})\ket{0} \, .
\end{equation}
In terms of  the $\mathcal{O}_{[\Delta, n, m]}(0,{\bf 0})$, we have
\begin{align}
[D,\mathcal{O}_{[\Delta, n, m]}(0,{\bf 0})]=i \Delta ~\mathcal{O}_{[\Delta, n, m]}(0,{\bf 0}) ~~~\rightarrow~~~ e^{i\lambda D}\mathcal{O}_{[\Delta, n, m]}(0,{\bf 0})e^{-i\lambda D}=e^{-\lambda\Delta} \mathcal{O}_{[\Delta, n, m]}(0,{\bf 0}) \, .
\end{align}
The factor of $i$ ensures the operator $\mathcal{O}_{[\Delta, n, m]}(0,{\bf 0})$ gets scaled by a real number $\exp(-\lambda\Delta)$ under the transformation generated by $D$. 

After some algebra, one can find the following relations, which follow directly from the \Sch algebra (see Table~\ref{weighttable}):
\begin{eqnarray}
DP_i \ket{\Delta, n,  m} &=& i(\Delta+1)P_i\ket{\Delta, n,  m}  \,, \\
DK_i \ket{\Delta, n, m} &=& i(\Delta-1)K_i\ket{\Delta, n,  m}  \,, \\
DH \ket{\Delta, n, m} &=& i(\Delta+2)H\ket{\Delta, n,  m}  \,, \\
DC \ket{\Delta, n,  m} &=& i(\Delta-2)C\ket{\Delta, n,  m}  \,,
\end{eqnarray}
%
or, written another way:
\begin{eqnarray}\label{start}
\left[D,[P_i, \mathcal{O}_{[\Delta, n, m]}(0,{\bf 0})]\right] \ket{0} &=& i(\Delta+1) [P_i, \mathcal{O}_{[\Delta, n, m]}(0,{\bf 0})]\ket{0}  \,, \\
\left[D,[K_i, \mathcal{O}_{[\Delta, n, m]}(0,{\bf 0})]\right] \ket{0} &=& i(\Delta-1) [K_i, \mathcal{O}_{[\Delta, n, m]}(0,{\bf 0})]\ket{0}  \,, \\
\left[D,[H, \mathcal{O}_{[\Delta, n, m]}(0,{\bf 0})]\right] \ket{0} &=& i(\Delta+2) [H, \mathcal{O}_{[\Delta, n, m]}(0,{\bf 0})]\ket{0} \, , \\
\left[D,[C, \mathcal{O}_{[\Delta, n, m]}(0,{\bf 0})]\right] \ket{0} &=& i(\Delta-2) [C, \mathcal{O}_{[\Delta, n, m]}(0,{\bf 0})]\ket{0}  \,.
\end{eqnarray}
Therefore, $P_i$ and $H$ act as lowering operators, and $K_i$ and $C$ act as raising operators, in analogy to the representation of $SU(2)$ or $SO(3)$.  If one assumes that the scaling dimension $\Delta$ of operators in this Hilbert space is bounded from below, then there will be a set operators $\mathcal{O}^P$ such that:
\begin{equation}
\label{primaries}
{[K_i, \mathcal{O}^P_{[\Delta, n , m]}(0, {\bf 0})]}\ket{0} = 0 \, , \hspace{0.25in} \text{and} \hspace{0.25in} [C, \mathcal{O}^P_{[\Delta, n , m]}(0, {\bf 0})]\ket{0} =0  \,.
\end{equation}
Such operators are called {\it primary operators}, which are associated with the highest-weight states.\footnote{These are actually states with lowest scaling dimension. Following the group theory literature, we call them the highest-weight state.}  Acting with $P_i$ or $H$ repeatedly on these primary operators produces a tower of operators, where $P_i$ and $H$ raise the scaling dimension by either one or two units.  Acting with $P_i$ and $H$ are associated with space and time derivatives:
\begin{eqnarray}
\label{derivatives}\label{end}
{[P_i, \mathcal{O}(t,{\bf x})]} \ket{0} = i \partial_i \mathcal{O}(t,{\bf x}) \ket{0}, \hspace{0.25in} {[H, \mathcal{O}(t,{\bf x})]} \ket{0}= - i \partial_t \mathcal{O}(t,{\bf x}) \ket{0} \, ,
\end{eqnarray}
which is true for any operator $\mathcal{O}$.

The \Sch algebra permits further categorization of different kinds of primary operators.  Beginning with the Jacobi identity for any operators $A$, $B$, and $C$:
\begin{equation}
[A,[B,C]] + [B,[C,A]] + [C,[A,B]] = 0\,,
\end{equation}
one can arrive at the following identities:
\begin{eqnarray}\label{pri}
[K_i, [P_j, \mathcal{O}^P_{[\Delta, n , m]}(0, {\bf 0})]] &=& - i n \delta_{ij} \mathcal{O}^P_{[\Delta, n , m]}(0, {\bf 0}) \, , \\
{[C, [P_i, \mathcal{O}^P_{[\Delta, n , m]}(0, {\bf 0})]] }&=& 0  \,. 
\end{eqnarray}
For $n\neq0$, $[P_i, \mathcal{O}^P_{[\Delta, 0, m]}]$ are not primary operators and are called \textit{descendants}. Thus, when $n\neq 0$, all descendants are total spatial or time derivatives of some primary operator.  The distinction between primary and descendants become obfuscated in the $n=0$ sector.  This is because if $n= 0$, then $[P_i, \mathcal{O}^P_{[\Delta, 0, m]}]$ is also a primary operator, as evident from Eq.~\eqref{pri}. In this scenario, one can reorganize the operators by being agnostic to the action of $K_i$, following Ref.~\cite{Pal:2018idc}. In particular, one categorize the operators into \textit{quasi-primaries} and its descendants, where the quasi-primaries satisfy $[C,\mathcal{O}(0, {\bf 0})] = 0$ without making any requirements for the value of $[K_i,\mathcal{O}(0, {\bf 0})] $, and the descendants are obtained by action of $H$ on quasi-primaries. This categorization exploits the SL(2,$\mathbb{R}$) subgroup generated by $H,D,C$ which applies for both the neutral and the charged sectors. 

An operator can be written as a total time derivative of another operator if and only if it is a descendant of a quasi-primary, which leaves out the possibility that a quasi-primary can be a total spatial derivative of some operator.  This necessitates further categorization:~we define $\mathcal{O}^{PA}_{[\Delta, n , m]}(0, {\bf 0})]$ to be \textit{quasi-primaries of type-A} if and only if satisfies
\begin{align}
[C,\mathcal{O}^{PA}_{[\Delta, n , m]}(0, {\bf 0})]=0\,,\\
\label{typeA}
\mathcal{O}^{PA}_{[\Delta, n , m]}(0, {\bf 0}) \neq[P_i,\mathcal{O}]\,,
\end{align}
for any operator $\mathcal{O}$.  \textit{Quasi-primaries of type-B} are those where Eq.~\eqref{typeA} does not hold.

\section{Constraints from Algebra and Unitarity bound}
\label{app3}

The \Sch algebra completely restricts the spacetime dependence of the two-point correlation function between two primary operators.  Consider two primary operators, $\mathcal{O}_1$ with scaling dimension $\Delta_1$ and number charge $n_1$ and a second, $\mathcal{O}_2$, with scaling dimension $\Delta_2 $ and number charge $n_2$.  Begin with the following expressions, which are explicitly zero, since all generators annihilate the vacuum:
\begin{eqnarray}
\label{zero1}
\bra{0} [K_i, \mathcal{O}_2(t, {\bf x})] \mathcal{O}_1(0, {\bf 0}) \ket{0} + \bra{0}\mathcal{O}_2(t, {\bf x}) [K_i, \mathcal{O}_1(0, {\bf 0})]  \ket{0} & =& 0  \,, \\
\bra{0} [C, \mathcal{O}_2(t, {\bf x})] \mathcal{O}_1(0, {\bf 0}) \ket{0} + \bra{0}\mathcal{O}_2(t, {\bf x}) [C, \mathcal{O}_1(0, {\bf 0})]  \ket{0} & =& 0 \, , \\
\bra{0} [D, \mathcal{O}_2(t, {\bf x})] \mathcal{O}_1(0, {\bf 0}) \ket{0} + \bra{0}\mathcal{O}_2(t, {\bf x}) [D, \mathcal{O}_1(0, {\bf 0})]  \ket{0} & =& 0 \, , \\
\label{zero4}
\bra{0} [N, \mathcal{O}_2(t, {\bf x})] \mathcal{O}_1(0, {\bf 0}) \ket{0} + \bra{0}\mathcal{O}_2(t, {\bf x}) [N, \mathcal{O}_1(0, {\bf 0})]  \ket{0} & =& 0 \, .
\end{eqnarray}
One can use $\mathcal{O}(t,{\bf x}) = e^{-i(Ht - P_j x_j)}\mathcal{O}(0,{\bf 0})e^{i(Ht - P_j x_j)}$ and Eq.~\eqref{derivatives} to generalize the relations in Eq.~\eqref{primaries} to an arbitrary point in spacetime:
%
\begin{eqnarray}
\label{primaryKxt}
{[K_i, \mathcal{O}^P_{[\Delta, n , m]}(t, {\bf x})]}\ket{0} &=& \left( nx_i  - it\partial_i \right)\mathcal{O}^P_{[\Delta, n , m]}(t, {\bf x}) \ket{0}  \,, \\
\label{primaryCxt}
{[C, \mathcal{O}^P_{[\Delta, n , m]}(t, {\bf x})] }\ket{0} &=& \left( -it\Delta + \frac{1}{2}x^2 n - it x_j \partial_j - i t^2 \partial_t \right)\mathcal{O}^P_{[\Delta, n , m]}(t, {\bf x}) \ket{0} \, , \\
\label{primaryDxt}
{[D, \mathcal{O}^P_{[\Delta, n , m]}(t, {\bf x})]}\ket{0} &=& i\left( \Delta + x_j \partial_j + 2t \partial_t \right)\mathcal{O}^P_{[\Delta, n , m]}(t, {\bf x}) \ket{0}  \, .
\end{eqnarray}
Inserting Eqs.~\eqref{primaryKxt} - \eqref{primaryDxt} in to Eqs.~\eqref{zero1} -~\eqref{zero4}:
%
\begin{align}
\bra{0} \left(n_2 x_i - it\partial_i)\mathcal{O}_2(t, {\bf x}\right) \mathcal{O}_1(0, {\bf 0}) \ket{0} & = 0\, , \\
\bra{0} \left(- i t \Delta_2 + \frac{1}{2} x^2 n_2 - itx_j \partial_j - it^2 \partial_t \right) \mathcal{O}_2(t, {\bf x}) \mathcal{O}_1(0, {\bf 0})\ket{0}   &= 0\, , \\
\bra{0} \left( x_j \partial_j + 2t \partial_t + \Delta_1 + \Delta_2 \right) \mathcal{O}_2(t, {\bf x}) \mathcal{O}_1(0, {\bf 0}) \ket{0}  &= 0\, , \\
(n_1 + n_2) \bra{0}\mathcal{O}_2(t, {\bf x}) \mathcal{O}_1(0, {\bf 0}) \ket{0}&  = 0\,.
\end{align}
This system of differential equations can be simultaneously solved to give a non-trivial result only if $\Delta_1 = \Delta_2 \equiv \Delta$ and $n_2 = - n_1 \equiv n$, which can be satisfied if $\mathcal{O}_1 = \mathcal{O}_2^\dagger \equiv \mathcal{O}^\dagger$, and if so \cite{Nishida:2007pj,Goldberger:2014hca}:
\begin{equation}
\label{2pcf}
\bra{0} \mathcal{O}(t, {\bf x} ) \mathcal{O}^\dagger(0, {\bf 0}) \ket{0} = \text{const} \cdot t^{-\Delta} \exp\left[\frac{-inx^2}{2t} \right]  \, .
\end{equation}
This is true whether $t$ is positive or negative, therefore the time-ordered product $\bra{0} T\mathcal{O}(t, {\bf x} ) \mathcal{O}^\dagger(0, {\bf 0}) \ket{0}$  has the same form.

There are additional constraints on operators that come from the requirement of unitarity. There are various ways to arrive at the unitarity bound \cite{Nishida:2007pj,Goldberger:2014hca,Pal:2018idc}. Here we will briefly sketch the method described in Ref.~\cite{Pal:2018idc}.
Let $\tau = it$, and assume the state $\ket{\psi(\tau,{\bf x})}$ can be associated with a local primary operator, with scaling dimension $\Delta$ and number charge $n$, acting on the vacuum: $\ket{\psi(\tau,{\bf x})} \equiv \mathcal{O}^\dagger(\tau,{\bf x}) \ket{0}$.  Then,  the requirement of unitarity can be written as:
\begin{equation}
\label{unitarity2}
\langle \psi(\tau,{\bf x}) | \psi(\tau,{\bf x}) \rangle \geq 0 \Leftrightarrow \lim_{\substack{\tau' \rightarrow -\tau \\ {\bf x'} \rightarrow {\bf x}}} \bra{0} \mathcal{O}(\tau', {\bf x'}) \mathcal{O}^\dagger(\tau, {\bf x}) \ket{0} \geq 0 \, .
\end{equation}
Because these are primary operators, the form of this two-point correlation function from Eq.~\eqref{2pcf} is
\begin{equation}
\label{2pcfuni}
\bra{0} \mathcal{O}(\tau', {\bf x'}) \mathcal{O}^\dagger(\tau, {\bf x}) \ket{0} = \text{const} \cdot (\tau' - \tau)^{-\Delta} \exp\left[ \frac{n({\bf x' - x})^2}{2(\tau' - \tau)} \right].
\end{equation}
The algebra does not constrain the overall constant in the above expression, so one is free to choose it such that
\begin{equation}
\label{constfreedom}
\text{const} \cdot ( -\tau)^{-\Delta}  \geq 0  \,,
\end{equation}
is always true.  Because Eq.~\eqref{unitarity2} is required for any state $\ket{\psi}$ in the Hilbert space, it must also hold for a state $\ket{\widetilde{\psi}}$ defined as any combination of partial (Euclidean) time and space derivatives acting on the original primary operator: $\ket{\widetilde{\psi}} \equiv (A \partial_\tau + B \partial_i + C \partial_i \partial_j + ...)\mathcal{O}^\dagger(\tau, {\bf x}) \ket{0}$, where $A$, $B$, $C$, ..., are constants.  Consider a particular state:
\begin{equation}
\label{psitilde}
\ket{\widetilde{\psi}} \equiv (\alpha \partial_\tau + \beta \partial_i \partial_i) \mathcal{O}^\dagger(\tau, {\bf x}) \ket{0}  \, ,
\end{equation}
where $\alpha$ and $\beta$ are real constants.  The requirement that $\langle \widetilde{\psi}(\tau,{\bf x}) | \widetilde{\psi}(\tau,{\bf x}) \rangle \geq 0$ then leads to the following inequality:
\begin{equation}
\lim_{\substack{\tau' \rightarrow -\tau \\ {\bf x'} \rightarrow {\bf x}}}( -\alpha \partial_{\tau'} + \beta \partial_{i'} \partial_{i'})( \alpha \partial_\tau + \beta \partial_i \partial_i) \bra{0} \mathcal{O}(\tau', {\bf x'}) \mathcal{O}^\dagger(\tau, {\bf x}) \ket{0} \geq 0 \, ,
\end{equation}
where the primes indicate that they only act on the primed spacetime variables.  Using Eqs.~\eqref{2pcfuni} and~\eqref{constfreedom}, this leads to the inequality:
\begin{equation}\label{unitaritycond}
\beta^2 n^2 (d^2 + 2 d)  + 2\alpha\beta  nd (\Delta+1) + \alpha^2 \Delta (\Delta + 1)  \geq 0  \, ,
\end{equation}
where $d$ is the number of spatial dimensions.  For a fixed value of $d$, the Eq.~\eqref{unitaritycond} implies that $\Delta\notin(-1,d/2)$. Similarly, considering a state of the form $(\alpha'\mathcal{O}^\dagger(\tau, {\bf x})+\partial_\tau\mathcal{O}^\dagger(\tau, {\bf x})) \ket{0}$, one can rule out $\Delta<0$~\cite{Pal:2018idc}. Combining these two bounds, we have a bound on $\Delta$, i.e.,  $\Delta \geq d/2$, which occurs when $\alpha/\beta = -2n$.  If $\Delta = d/2$, then the state $\ket{\widetilde{\psi}}$ is a null state with zero norm, where, plugging these values for $\alpha$ and $\beta$ back into Eq.~\eqref{psitilde}, and setting $\Delta = d/2$:
\begin{equation}
\left(\partial_\tau - \frac{\partial_i^2}{2n} \right) \ket{{\psi}(\tau, {\bf x})} = 0  \, .
\end{equation}
Identifying $\tau = it$ and $-n$ (the charge of $\mathcal{O}^\dagger$ is $-n>0$) as the mass of the particle, we have
\begin{align}
\left(-i\partial_t + \frac{\partial_i^2}{2m} \right)\mathcal{O}^\dagger \ket{0} &=0\,,\ \left(i\partial_t+\frac{\partial_i^2}{2m}\right)\mathcal{O}\ket{0}=0\,.
\end{align}
This is the \Sch equation.  Because this is an example of a classical equation of motion for a well-defined quantum field theory, the bound $\Delta  \geq d/2$ is therefore the strongest bound.\footnote{To prove this, assume there is a better (or as good) lower bound $\Delta^*$ such that $\Delta \geq \Delta^* \geq d/2$.  There exists an example of a well-defined quantum field theory with $\Delta = d/2$, therefore $\Delta = \Delta^*$. }

\bibliographystyle{JHEP}

\bibliography{bib}{}

\providecommand{\href}[2]{#2}\begingroup\raggedright\begin{thebibliography}{10}

\bibitem{Politzer:1980me}
H.~D. Politzer, {\it {Power Corrections at Short Distances}},  {\em Nucl.
  Phys.} {\bf B172} (1980) 349--382.

\bibitem{Georgi:1991ch}
H.~Georgi, {\it {On-shell effective field theory}},  {\em Nucl. Phys.} {\bf
  B361} (1991) 339--350.

\bibitem{Kinoshita:1995mt}
T.~Kinoshita and M.~Nio, {\it {Radiative corrections to the muonium hyperfine
  structure. 1. The alpha**2 (Z-alpha) correction}},  {\em Phys. Rev.} {\bf
  D53} (1996) 4909--4929, [\href{http://arxiv.org/abs/hep-ph/9512327}{{\tt
  hep-ph/9512327}}].

\bibitem{Hill:2012rh}
R.~J. Hill, G.~Lee, G.~Paz, and M.~P. Solon, {\it {NRQED Lagrangian at order
  $1/M^4$}},  {\em Phys. Rev.} {\bf D87} (2013) 053017,
  [\href{http://arxiv.org/abs/1212.4508}{{\tt arXiv:1212.4508}}].

\bibitem{Manohar:1997qy}
A.~V. Manohar, {\it {The HQET / NRQCD Lagrangian to order alpha / m-3}},  {\em
  Phys. Rev.} {\bf D56} (1997) 230--237,
  [\href{http://arxiv.org/abs/hep-ph/9701294}{{\tt hep-ph/9701294}}].

\bibitem{Kobach:2017xkw}
A.~Kobach and S.~Pal, {\it {Hilbert Series and Operator Basis for NRQED and
  NRQCD/HQET}},  {\em Phys. Lett.} {\bf B772} (2017) 225--231,
  [\href{http://arxiv.org/abs/1704.00008}{{\tt arXiv:1704.00008}}].

\bibitem{Gunawardana:2017zix}
A.~Gunawardana and G.~Paz, {\it {On HQET and NRQCD Operators of Dimension 8 and
  Above}},  {\em JHEP} {\bf 07} (2017) 137,
  [\href{http://arxiv.org/abs/1702.08904}{{\tt arXiv:1702.08904}}].

\bibitem{Henning:2015daa}
B.~Henning, X.~Lu, T.~Melia, and H.~Murayama, {\it {Hilbert series and operator
  bases with derivatives in effective field theories}},  {\em Commun. Math.
  Phys.} {\bf 347} (2016), no.~2 363--388,
  [\href{http://arxiv.org/abs/1507.07240}{{\tt arXiv:1507.07240}}].

\bibitem{Henning:2015alf}
B.~Henning, X.~Lu, T.~Melia, and H.~Murayama, {\it {2, 84, 30, 993, 560, 15456,
  11962, 261485, ...: Higher dimension operators in the SM EFT}},  {\em JHEP}
  {\bf 08} (2017) 016, [\href{http://arxiv.org/abs/1512.03433}{{\tt
  arXiv:1512.03433}}].

\bibitem{Henning:2017fpj}
B.~Henning, X.~Lu, T.~Melia, and H.~Murayama, {\it {Operator bases,
  $S$-matrices, and their partition functions}},  {\em JHEP} {\bf 10} (2017)
  199, [\href{http://arxiv.org/abs/1706.08520}{{\tt arXiv:1706.08520}}].

\bibitem{Lehman:2015via}
L.~Lehman and A.~Martin, {\it {Hilbert Series for Constructing Lagrangians:
  expanding the phenomenologist's toolbox}},  {\em Phys. Rev.} {\bf D91} (2015)
  105014, [\href{http://arxiv.org/abs/1503.07537}{{\tt arXiv:1503.07537}}].

\bibitem{Luke:1992cs}
M.~E. Luke and A.~V. Manohar, {\it {Reparametrization invariance constraints on
  heavy particle effective field theories}},  {\em Phys. Lett.} {\bf B286}
  (1992) 348--354, [\href{http://arxiv.org/abs/hep-ph/9205228}{{\tt
  hep-ph/9205228}}].

\bibitem{Heinonen:2012km}
J.~Heinonen, R.~J. Hill, and M.~P. Solon, {\it {Lorentz invariance in heavy
  particle effective theories}},  {\em Phys. Rev.} {\bf D86} (2012) 094020,
  [\href{http://arxiv.org/abs/1208.0601}{{\tt arXiv:1208.0601}}].

\bibitem{Barabanschikov:2005ri}
A.~Barabanschikov, L.~Grant, L.~L. Huang, and S.~Raju, {\it {The Spectrum of
  Yang Mills on a sphere}},  {\em JHEP} {\bf 01} (2006) 160,
  [\href{http://arxiv.org/abs/hep-th/0501063}{{\tt hep-th/0501063}}].

\bibitem{Nishida:2007pj}
Y.~Nishida and D.~T. Son, {\it {Nonrelativistic conformal field theories}},
  {\em Phys. Rev.} {\bf D76} (2007) 086004,
  [\href{http://arxiv.org/abs/0706.3746}{{\tt arXiv:0706.3746}}].

\bibitem{Goldberger:2014hca}
W.~D. Goldberger, Z.~U. Khandker, and S.~Prabhu, {\it {OPE convergence in
  non-relativistic conformal field theories}},  {\em JHEP} {\bf 12} (2015) 048,
  [\href{http://arxiv.org/abs/1412.8507}{{\tt arXiv:1412.8507}}].

\bibitem{Pal:2018idc}
S.~Pal, {\it {Unitarity and universality in nonrelativistic conformal field
  theory}},  {\em Phys. Rev.} {\bf D97} (2018), no.~10 105031,
  [\href{http://arxiv.org/abs/1802.02262}{{\tt arXiv:1802.02262}}].

\bibitem{Rychkov:2015naa}
S.~Rychkov and Z.~M. Tan, {\it {The $\epsilon$-expansion from conformal field
  theory}},  {\em J. Phys.} {\bf A48} (2015), no.~29 29FT01,
  [\href{http://arxiv.org/abs/1505.00963}{{\tt arXiv:1505.00963}}].

\bibitem{Zwierlein:2004zz}
M.~W. Zwierlein, C.~A. Stan, C.~H. Schunck, S.~M.~F. Raupach, A.~J. Kerman, and
  W.~Ketterle, {\it {Condensation of Pairs of Fermionic Atoms near a Feshbach
  Resonance}},  {\em Phys. Rev. Lett.} {\bf 92} (2004) 120403.

\bibitem{Regal:2004zza}
C.~A. Regal, M.~Greiner, and D.~S. Jin, {\it {Observation of Resonance
  Condensation of Fermionic Atom Pairs}},  {\em Phys. Rev. Lett.} {\bf 92}
  (2004) 040403.

\bibitem{Kaplan:1998we}
D.~B. Kaplan, M.~J. Savage, and M.~B. Wise, {\it {Two nucleon systems from
  effective field theory}},  {\em Nucl. Phys.} {\bf B534} (1998) 329--355,
  [\href{http://arxiv.org/abs/nucl-th/9802075}{{\tt nucl-th/9802075}}].

\bibitem{Kaplan:1998tg}
D.~B. Kaplan, M.~J. Savage, and M.~B. Wise, {\it {A New expansion for
  nucleon-nucleon interactions}},  {\em Phys. Lett.} {\bf B424} (1998)
  390--396, [\href{http://arxiv.org/abs/nucl-th/9801034}{{\tt
  nucl-th/9801034}}].

\bibitem{Arav:2016xjc}
I.~Arav, S.~Chapman, and Y.~Oz, {\it {Non-Relativistic Scale Anomalies}},  {\em
  JHEP} {\bf 06} (2016) 158, [\href{http://arxiv.org/abs/1601.06795}{{\tt
  arXiv:1601.06795}}].

\bibitem{Pal:2016rpz}
S.~Pal and B.~Grinstein, {\it {Weyl Consistency Conditions in Non-Relativistic
  Quantum Field Theory}},  {\em JHEP} {\bf 12} (2016) 012,
  [\href{http://arxiv.org/abs/1605.02748}{{\tt arXiv:1605.02748}}].

\bibitem{Auzzi:2016lrq}
R.~Auzzi, S.~Baiguera, F.~Filippini, and G.~Nardelli, {\it {On Newton-Cartan
  local renormalization group and anomalies}},  {\em JHEP} {\bf 11} (2016) 163,
  [\href{http://arxiv.org/abs/1610.00123}{{\tt arXiv:1610.00123}}].

\bibitem{levy1971galilei}
J.~M. L{\'e}vy-Leblond, {\em Galilei group and Galilean invariance}.
\newblock Academic Press, 1971.

\bibitem{Niederer:1972zz}
U.~Niederer, {\it {The maximal kinematical invariance group of the free
  Schrodinger equation.}},  {\em Helv. Phys. Acta} {\bf 45} (1972) 802--810.

\bibitem{Henkel:1993sg}
M.~Henkel, {\it {Schrodinger invariance in strongly anisotropic critical
  systems}},  {\em J. Statist. Phys.} {\bf 75} (1994) 1023--1061,
  [\href{http://arxiv.org/abs/hep-th/9310081}{{\tt hep-th/9310081}}].

\bibitem{Mehen:1999nd}
T.~Mehen, I.~W. Stewart, and M.~B. Wise, {\it {Conformal invariance for
  nonrelativistic field theory}},  {\em Phys. Lett.} {\bf B474} (2000)
  145--152, [\href{http://arxiv.org/abs/hep-th/9910025}{{\tt hep-th/9910025}}].

\bibitem{Henkel:2003pu}
M.~Henkel and J.~Unterberger, {\it {Schrodinger invariance and space-time
  symmetries}},  {\em Nucl. Phys.} {\bf B660} (2003) 407--435,
  [\href{http://arxiv.org/abs/hep-th/0302187}{{\tt hep-th/0302187}}].

\bibitem{Golkar:2014mwa}
S.~Golkar and D.~T. Son, {\it {Operator Product Expansion and Conservation Laws
  in Non-Relativistic Conformal Field Theories}},  {\em JHEP} {\bf 12} (2014)
  063, [\href{http://arxiv.org/abs/1408.3629}{{\tt arXiv:1408.3629}}].

\end{thebibliography}\endgroup

\end{document}